\documentclass[aps,showpacs,twocolumn]{revtex4}
\usepackage{amsfonts}
\usepackage{amsmath}
\usepackage{amssymb}
\usepackage{graphicx}
\usepackage{bm}

\input{tcilatex}

\begin{document}
\title{Electric-field induced capillary interaction of charged particles at\
a polar interface}
\author{Lionel Foret and Alois W\"{u}rger}
\address{CPMOH\thanks{%
Unit\'{e} Mixte de Recherche CNRS 5798}, Universit\'{e} Bordeaux 1, 351 cours de la Lib\'{e}ration, 33405 Talence, France}

\begin{abstract}
We study the electric-field induced capillary interaction of charged
particles at a polar interface. The algebraic tails of the electrostatic
pressure of each charge results in a deformation of the interface $u\sim
\rho ^{-4}$. The resulting capillary interaction\ is repulsive and varies as 
$\rho ^{-6}$ with the particle distance. As a consequence, electric-field
induced capillary forces cannot be at the origin of the secondary minimum
observed recently for charged PMMA particles at on oil-water interface.
\end{abstract}

\maketitle

Electrostatic forces operating on charged particles determine to a large
extent the properties of emulsions and foams, and stabilize colloidal
aggregates \cite{Tak99,Ave93}. Charged particles at interfaces or in thin
films form 2D hexagonal crystals \cite{Pie80,Pie83,Ave00,Ave02,Nik02}, while
for bulk colloidal suspensions several 3D crystal phases have been reported 
\cite{Yet03}. At a typical lattice spacing of a few microns, the
interparticle forces are of the order of picoNewton.

A charge at a polar interface and its counterion cloud carry a finite dipole
moment perpendicular to the interface (Cf. Fig. 1.) The resulting
electrostatic force acting on two neighboring particles is repulsive and
varies with the inverse fourth power of their distance \cite%
{Pie80,Sti61,Hur85,Net99,For03}. This law has been confirmed experimentally
for charged polystyrene particles \cite{Ave02}.

The pair potential of PMMA particles at an oil-water interface has been
measured by recording the trajectories and evaluating distance correlations.
Besides the dipolar repulsion at short distances, an attractive force has
been found at larger distances, and a minimum in the potential energy has
been shown to occur at about 5.7 ${\rm \mu }$m \cite{Nik02}. As a possible
explanation, these authors propose that the deformation of the interface by
the particles gives rise to a capillary interaction that varies
logarithmically with their distance \cite{Nik02}.

Quite generally, a capillary or elastic interaction is obtained when
equilibrating an external force acting on the particles with the surface
tension or the bending rigidity. A variety of such models have been studied,
both for spherical and anisotropic defects \cite%
{Par96,Bou98,Tur99,Sch00,Dan01,Mar02,Sta00}; these works deal with forces
that act on the particle only, such as gravity. A somewhat different pattern
arises for forces that operate both on the particle and on the surrounding
interface. For example, a polymer grafted on a membrane exerts a force at
the point of attachment and an opposite entropic pressure on the interface 
\cite{Bic00}.

In this Letter we study the deformation of a polar interface due to a
charged colloidal particle, and we derive the resulting capillary
interaction. The present work is confined to the case most relevant for
micron size colloidal particles, where the distance is much larger than the
Debye length. After a reminder of the free energy of a deformation field, we
calculate the electrostatic pressure profile exerted by a charge and the
associated counterions (as shown schematically in Fig. 1.) The
electric-field induced capillary interaction is compared to experimental
findings and to the case of $\delta $-force that was proposed in \cite{Nik02}%
.

\begin{figure}
\includegraphics[width=\columnwidth]{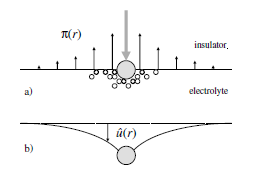}
\caption{a) Pressure profile $\hat{\protect\pi}(r)$ induced on the interface
by a charged particle (filled grey circle); open circles are the
counterions. b) The corresponding deformation field $\hat{u}(r)$.}
\end{figure}

We consider particles of charge $Q$ trapped at an electrolyte-insulator
interface. In the absence of forces, the interface $S$ is flat and its
energy reads $\gamma S$, where $\gamma $ is the surface tension. The charged
particles and their counterions exert on the interface a pressure\ that is
given by the normal component of the stress tensor and comprises entropic
and electrostrictive contributions, \ 
\begin{equation}
\pi ({\bf r})=k_{B}T(n_{+}+n_{-})+{\cal T}_{zz}^{{\rm ins}}-{\cal T}_{zz}^{%
{\rm el}}.  \label{eq2}
\end{equation}
The first term is the entropic pressure of the excess density of positive
and negative ions, $n_{\pm }=n_{s}\left( e^{\mp e\phi (r)/k_{B}T}-1\right) $%
, induced by $N$ surface charges. Electroneutrality requires $e\int
dV(n_{+}-n_{-})=-NQ$. The remaining terms involve the Maxwell tensor 
\begin{equation}
{\cal T}_{ij}=\varepsilon \left( E_{i}E_{j}-\frac{1}{2}{\bf E}^{2}\delta
_{ij}\right) ,  \label{eq4}
\end{equation}
evaluated at the electrolyte and insulating side of the interface. Note that
both the dielectric constant $\varepsilon $ and the normal component $E_{z}$
of the electric field vector are discontinuous across the interface. The
corresponding jump of the normal component of the Mawell tensor describes an
electrostrictive force that arises from the difference in the electric field
density in the two media.

As a consequence of the inhomogeneous pressure, the interface is no longer
flat; the deformation $u({\bf r})$ increases the total surface and thus the
surface energy by the amount 
\[
\gamma \int dS\left( \sqrt{1+\left( {\bf \nabla }u\right) ^{2}}-1\right) , 
\]
where ${\bf \nabla }$ is the 2D gradient in the interface plane. In the
following, ${\bf r}$ denotes the in-plane coordinates, and $z$ that normal
to the interface. In the case of weak deformations, $|{\bf \nabla }u|\ll 1$,
we may expand the square root and retain the leading term only. Then the
free energy functional reads 
\begin{equation}
f[u({\bf r})]=\frac{\gamma }{2}\int dS\left( {\bf \nabla }u\right) ^{2}-\int
dS\pi ({\bf r})u({\bf r}).  \label{eq6}
\end{equation}
The actual deformation is determined by minimizing this functional.
Linearizing the deformation about the equilibrium value $u({\bf r})$ and
integrating the first term by parts, one finds that the mimimum free energy $%
f$ occurs for the deformation satisfying the Young-Laplace equation, 
\begin{equation}
\gamma \nabla ^{2}u({\bf r})+\pi ({\bf r})=0,  \label{eq9}
\end{equation}
and reads 
\begin{equation}
f=-\frac{\gamma }{2}\int dS\left( {\bf \nabla }u\right) ^{2}.  \label{eq18}
\end{equation}

As the most severe approximation of the present paper, we assume that the
total pressure $\pi ({\bf r})$ is the linear superposition 
\begin{equation}
\pi ({\bf r})=\sum_{\alpha }\hat{\pi}({\bf r-r}_{\alpha }),  \label{eq19}
\end{equation}
where $\hat{\pi}({\bf r})$ is the pressure profile of a single charge at the
origin. In Debye-H\"{u}ckel approximation, the electrostatic potential is a
linear superposition of single-particle terms $\phi =\sum_{\alpha }\varphi
_{\alpha }$, and the total pressure is quadratic in the potential and its
derivatives, $\pi (r)\sim \phi ^{2}$. Besides the diagonal terms $\hat{\pi}%
_{\alpha }\sim \varphi _{\alpha }^{2}$, it comprises cross-terms $\varphi
_{\alpha }\varphi _{\beta }$ that are significantly smaller and thus have
been neglected in (\ref{eq19}).

\begin{figure}
\includegraphics[width=\columnwidth]{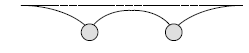}
\caption{The capillary interaction arises from the superposition of the
deformation fields of nearby particles.}
\end{figure}

Because of the linearity of (\ref{eq9}), a similar relation holds true for
the deformation 
\[
u({\bf r})=\sum_{\alpha }\hat{u}({\bf r-r}_{\alpha }). 
\]
It is then sufficient to consider two particles at positions $\pm \frac{1}{2}%
{\bf \rho }$. Discarding terms that do not depend on their distance $\rho $,
one finds immediatly the pair potential 
\begin{equation}
\Delta f(\rho )=-\gamma \int d^{2}r{\bf \nabla }\hat{u}\left( {\bf r}+\frac{1%
}{2}{\bf \rho }\right) \cdot {\bf \nabla }\hat{u}\left( {\bf r}-\frac{1}{2}%
{\bf \rho }\right) .  \label{eq26}
\end{equation}

The single-charge pressure $\hat{\pi}({\bf r})$ comprises the entropic
pressure of the excess soluted ion density and the electric field energy. It
is obvious from Fig.\ 1 that the force on the charge is directed towards the
electrolyte, whereas the electrostrictive force on the surrounding interface
is in the opposite direction. It is essential to note that the total force
vanishes, i.e., 
\begin{equation}
\int dS\hat{\pi}\left( r\right) =0.  \label{eq20}
\end{equation}
In view of (\ref{eq18}) we need to calculate the gradient of the deformation
field. This is achieved most easily by applying Gauss' theorem on (\ref{eq9}%
), 
\[
\gamma \oint_{\partial D}ds{\bf n}\cdot {\bf \nabla }\hat{u}=-\int_{D}dS\hat{%
\pi}\left( r\right) , 
\]
that relates the pressure integrated over the area $D$ to the \ ``flux'' of
the gradient field through the boundary $\partial D$. For a disk centered at
the origin and polar coordinates $r$ and $\theta $, we have $ds=rd\theta $
and ${\bf n=r}/r$. Using the condition (\ref{eq20}) and the fact that $\hat{u%
}$ is isotropic, the integral over the disk is replaced by that over the
infinite space outside, 
\begin{equation}
{\bf \nabla }\hat{u}(r)=\frac{1}{\gamma }\frac{{\bf r}}{r^{2}}%
\int_{r}^{\infty }dr^{\prime }r^{\prime }\hat{\pi}(r^{\prime }).
\label{eq24}
\end{equation}
Thus we have expressed the interaction free energy $\Delta f(\rho )$ by
integrals over the single-charge electrostatic pressure $\hat{\pi}(r)$ that
remains to be determined.

Starting from the screened potential $\varphi (r,z)$ of a single interfacial
charge at the origin, we evaluate the pressure profile according to Eqs. (%
\ref{eq2}) and (\ref{eq4}). Expanding the ion densities to quadratic order, $%
n_{+}+n_{-}=n_{s}\left( e\varphi /k_{B}T\right) ^{2}$, and\ noting $\kappa
^{2}=(2n_{s}e^{2}/\varepsilon _{{\rm el}}k_{B}T),$ we find the entropic
pressure 
\[
k_{B}T(n_{+}+n_{-})=\frac{\varepsilon _{{\rm el}}}{2}\kappa ^{2}\varphi
(r,z)^{2}. 
\]
The relevant component ${\cal T}_{zz}=\frac{\varepsilon }{2}%
(E_{z}^{2}-E_{r}^{2})$ of the Maxwell tensor depends on both normal and
in-plane derivatives of the potential. The latter one, $E_{r}=\partial
_{r}\varphi $, being continuous at the interface, the electrostrictive force
simplifies to 
\[
{\cal T}_{zz}^{{\rm ins}}-{\cal T}_{zz}^{{\rm el}}=\frac{\varepsilon _{{\rm %
el}}-\varepsilon _{{\rm ins}}}{2}E_{r}^{2}-\left. \frac{\varepsilon _{{\rm el%
}}}{2}E_{z}^{2}\right| _{{\rm el}}+\left. \frac{\varepsilon _{{\rm ins}}}{2}%
E_{z}^{2}\right| _{{\rm ins}} 
\]

In Debye-H\"{u}ckel approximation (DHA), the 2D Fourier transform of $%
\varphi $ is known, and explicit forms of the potential at the interface can
be given in the domains separated by the Bjerrum length $\ell _{B}$ and the
Debye screening length $\kappa ^{-1}$. Here we confine ourselves to the case
of large distances well beyond $\kappa ^{-1}$ and the particle size $a$,
where the potential in the electrolyte reads \cite{Sti61,Hur85,Net99,For03} 
\[
\varphi _{{\rm el}}(r,z)=(Q\varepsilon _{{\rm ins}}/2\pi \kappa
^{2}\varepsilon _{{\rm el}}^{2})e^{-\kappa z}r^{-3}\;\;\;\;(r\gg \kappa
^{-1}). 
\]
(High charge densities may result in a renormalized value for $Q$.) The
in-plane electric field $E_{r}\sim r^{-4}$ is negligible at large $r$; for
the normal component one finds $E_{z}|_{{\rm el}}=\partial _{z}\varphi _{%
{\rm el}}=-\kappa \varphi _{{\rm el}}$. Its value at the insulating side of
the interface is obtained from the continuity condition $\varepsilon _{{\rm %
ins}}E_{z}|_{{\rm ins}}=\varepsilon _{{\rm el}}E_{z}|_{{\rm el}}$. Thus the
first term in ${\cal T}_{zz}^{{\rm ins}}-{\cal T}_{zz}^{{\rm el}}$ cancels
the entropic pressure, the second one is negligible, and the third one gives 
\begin{equation}
\hat{\pi}(r)=\frac{\varepsilon _{{\rm ins}}}{2}\left( \frac{\varepsilon _{%
{\rm el}}}{\varepsilon _{{\rm ins}}}\kappa \varphi _{{\rm el}}(r,0)\right)
^{2}.  \label{eq28}
\end{equation}
Now it is straightforward to calculate the gradient of the interface
deformation, 
\begin{equation}
{\bf \nabla }\hat{u}(r)=\frac{1}{32\pi ^{2}}\frac{\varepsilon _{{\rm ins}%
}Q^{2}}{\gamma \varepsilon _{{\rm el}}^{2}\kappa ^{2}}\frac{{\bf r}}{r^{6}},
\label{eq32}
\end{equation}
This remarkable result, ${\bf \nabla }\hat{u}\sim r^{-5}$, relies only on
the asymptotic form of the potential $\varphi _{{\rm el}}\sim r^{-3}$ and
the fact that the net force (\ref{eq20}) on the interface vanishes. As a
consequence, the deformation field is negative and varies as $\hat{u}\sim
(-1/r^{4})$.

Now we discuss the behavior of $\hat{\pi}(r)$\ and ${\bf \nabla }\hat{u}(r)$%
\ at small distances.\ The singularities for $r\rightarrow 0$ are clearly
unphysical and would cause the deformation free energy $\Delta f$ to
diverge. The above screened potential $\varphi _{{\rm el}}$ and
electrostatic pressure $\hat{\pi}$\ are valid at large distances $r\gg
\kappa ^{-1},a$.\ At intermediate distances $a<r<\kappa ^{-1}$ screening is
irrelevant; evaluating the stress tensor with the bare electrostatic
potential $\varphi _{{\rm el}}\sim r^{-1}$, one finds a modified power law $%
\hat{\pi}(r)\sim r^{-4}$ \cite{For032}. For micrometer size charged
colloids, however, the particle radius $a$ provides the most relevant
physical cutoff.\ For this reason, it will be referred to in the remainder
of this paper, although one should keep in mind that the actual situation
may be more complex, especially for small molecules.

Thus the power laws of both $\hat{\pi}(r)$\ and ${\bf \nabla }\hat{u}(r)$\
cease to be valid at distances smaller than $a$.\ Moreover, Eq. (\ref{eq20})
requires a strong negative force operating on the particle at $r=0$. An
important conclusion can be drawn directly from the relation (\ref{eq20}): $%
{\bf \nabla }\hat{u}$ is positive everywhere and tends towards zero at $r=0$.

In order to regularize the surface integral in Eq. (\ref{eq26}), we
explicitly introduce a cut-off $a$ and replace (\ref{eq32}) with 
\begin{equation}
{\bf \nabla }\hat{u}(r)=\frac{1}{32\pi ^{2}}\frac{\varepsilon _{{\rm ins}%
}Q^{2}}{\gamma \varepsilon _{{\rm el}}^{2}\kappa ^{2}}\frac{{\bf r}}{%
(r^{2}+a^{2})^{3}}.
\end{equation}
Although the precise form of this cut-off function is somewhat arbitrary, it
satisfies the limit ${\bf \nabla }\hat{u}=0$ for $r\rightarrow 0$ as imposed
by Eq. (\ref{eq24}). For $r\gg a$, it shows the long-range behavior that has
been obtained rigorously from (\ref{eq20}) and (\ref{eq28}).

Inserting ${\bf \nabla }\hat{u}({\bf r}\pm \frac{1}{2}{\bf \rho })$ in (\ref%
{eq26}), one finds that the capillary interaction 
\[
\Delta f(\rho )=-(\varepsilon _{{\rm ins}}Q^{2}/32\pi ^{2}\gamma \varepsilon
_{{\rm el}}^{2}\kappa ^{2})^{2}I 
\]
depends on particle distance and size through the integral 
\[
I=\int_{0}^{\infty }drr\int_{0}^{2\pi }d\theta \frac{r^{2}-\frac{1}{4}\rho
^{2}}{\left( \left( r^{2}+\frac{1}{4}\rho ^{2}+a^{2}\right) ^{2}-r^{2}\rho
^{2}\cos ^{2}\theta \right) ^{3}}. 
\]
This definite integral can be performed analytically. Since in all
applications, the particle distance $\rho $ exceeds significantly the size $%
a $, we retain only the leading term in powers of $(a/\rho )$ and thus have 
\[
I=-\frac{2\pi }{a^{2}\rho ^{6}}\left( 1+{\cal O}\left( a^{2}/\rho
^{2}\right) \right) . 
\]
The variation with the particle radius $\sim a^{-2}$ does not depend on the
precise form of the cut-off function chosen above. Thus we obtain the
capillary interaction

\begin{equation}
\Delta f(\rho )=\frac{1}{2^{9}\pi ^{3}}\frac{\varepsilon _{{\rm ins}%
}^{2}Q^{4}}{\varepsilon _{{\rm el}}^{4}\kappa ^{4}}\frac{1}{\gamma a^{2}\rho
^{6}}  \label{eq36}
\end{equation}
that is repulsive and varies with the inverse sixth power of the distance.
Note that $\Delta f$ depends on the particle charge, its size, and the Debye
length.

We compare our result for the capillary interaction with experimental
findings and previous work. In Ref. \cite{Nik02}, a minimum in the
interaction potential of PMMA particles was reported to occur at a distance
of 5.7 microns. As a possible explanation, these authors considered the
competing electrostatic and capillary interactions. Interfacial charges are
subject to the well-known repulsive potential 
\[
V=\frac{\varepsilon _{{\rm ins}}}{2\pi \varepsilon _{{\rm el}}^{2}}\frac{%
p^{2}}{\rho ^{3}}, 
\]
where $p$ is dipole moment formed by the charge and its screening cloud \cite%
{Hur85}. On the other hand, assuming an electric-field induced $\delta $%
-force $\hat{\pi}({\bf r})=-\pi _{0}\delta ({\bf r})$, one obtains an
attractive capillary interaction $\Delta f_{\delta }(\rho )=(\pi
_{0}^{2}/2\pi )\ln \rho $. The sum of $V$ and \ $\Delta f_{\delta }$ shows a
minimum at finite distance that has been considered in Ref.\ \cite{Nik02}.
Note, however, that this force does not satisfy the condition (\ref{eq20}).

As shown schematically in Fig. 1 and discussed below (\ref{eq32}), the
electrostatic pressure profile induced by a screened surface charge
comprises a long-range contribution and a $\delta $-force, 
\[
\hat{\pi}({\bf r})=g(r)r^{-6}-\pi _{0}\delta ({\bf r}), 
\]
with a cut-off function $g$ that is constant for $r\gg a$, vanishes at the
origin, and satisfies Eq. (\ref{eq20}). By now it should be clear that the
long-range contribution to the pressure completely changes the capillary
forces: While that of a pure $\delta $-force is attractive and depends
logarithmically on distance, the capillary interaction of interfacial
charges is repulsive and varies as $\Delta f(\rho )\sim \rho ^{-6}$. We
conclude that the mimimum in the interaction potential reported in \cite%
{Nik02} cannot arise from $V+\Delta f$.

Finally we compare the relative magnitude of the forces arising from $V$ and 
$\Delta f$. If the particle's charge $Q$ is located at the interface, the
dipole moment reads $p=Q/\kappa $, and the capillary interaction can be
expressed in terms of the dipolar interaction $V$ and the surface energy $%
\gamma a^{2}$, 
\begin{equation}
\Delta f(\rho )=\frac{V^{2}}{2^{7}\pi \gamma a^{2}}.
\end{equation}
The total force $-\partial _{\rho }(V+\Delta f)$ is always repulsive. A
change in the exponent occurs where $\partial _{\rho }V=\partial _{\rho
}\Delta f$, defining the cross-over distance 
\[
\rho ^{\ast }=\frac{1}{4}\left( \frac{\varepsilon _{{\rm ins}}}{\pi
^{2}\varepsilon _{{\rm el}}^{2}}\frac{p^{2}}{\gamma a^{2}}\right) ^{1/3}, 
\]
for $\rho <\rho ^{\ast }$, the capillary repulsion $-\partial _{\rho }\Delta
f\sim \rho ^{-7}$ dominates, whereas at larger distances the dipolar force $%
-\partial _{\rho }V\sim \rho ^{-4}$\ takes over. For micron size particles
at an oil-water interface with typical parameters, one finds that the
cross-over distance hardly attains the micrometer range and thus is of
little relevance. Quite a different situation arises if a significant
fraction of the charge is located at the side of the particle immersed in
the insulator. For micron size particles, the electrostatic potential $%
\varphi $ is enhanced by a factor of the order $(\kappa a)^{2}$ \cite{Ave02}%
, and the capillary interaction increases by a factor $(\kappa a)^{4}$. The
modified dipole moment reads $p\sim Qa$\ instead of $Q/\kappa $, resulting
in a larger cross-over distance $\rho ^{\ast }$. Note, however, that at
distances much shorter than $\rho ^{\ast }$, the gradient of the deformation 
$|{\bf \nabla }\hat{u}|$ is no longer small, requiring to go beyond the
quadratic approximation for the surface energy in (\ref{eq6}).

In summary, we have studied the capillary interaction induced by the
electric field of charged particles at an electrolyte-insulator interface.
The electrostatic pressure of a single charge consists of a force acting on
the particle and the opposite electrostrictive force on the interface that
varies as $\rho ^{-6}$ with the lateral distance. The resulting deformation, 
$u\sim \rho ^{-4}$, induces a capillary interaction that is repulsive and
obeys a power law $\Delta f\sim \rho ^{-6}$. As a consequence, this
capillary force cannot be at the origin of the secondary minimum observed
recently for charged PMMA particles at an oil-water interface \cite{Nik02}.

$Note\;added$. After submission of this Letter, a comment by Megens and
Aizenberg on the paper by Nikolaides et al. \cite{Nik02} appeared in Nature 
\cite{Meg03}. In agreement with our discussion and our Eq. (8), Megens and
Aizenberg point out that there is no net force on the interface and thus
invalidate the attractive logarithmic potential proposed in Ref. \cite{Nik02}%
.

\end{document}